\documentclass[prd,twocolumn,superscriptaddress,nofootinbib,showpacs,noshowkeys]{revtex4} 

\usepackage{graphicx}
\usepackage{psfrag}

\begin{document}

\title{Exclusion regions and their power.}

\author{L.~Fleysher}
\email{lazar.fleysher@physics.nyu.edu}
\author{R.~Fleysher}
\email{roman.fleysher@physics.nyu.edu}
\affiliation{Department of Physics, New York University, New York,
                New York 10003}

\author{T.~J.~Haines}
\email{haines@lanl.gov}
\affiliation{Los Alamos National Laboratory, Los Alamos,
             New Mexico 87545}
\author{A.~I.~Mincer}
\email{allen.mincer@nyu.edu}
\author{P.~Nemethy}
\email{peter.nemethy@nyu.edu}
\affiliation{Department of Physics, New York University, New York,
                New York 10003}

\date{August 17, 2003}

\begin{abstract}

The question of exclusion region construction in new phenomenon
searches has been causing considerable discussions for many years and
yet no clear mathematical definition of the problem has been stated so
far. In this paper we formulate the problem in mathematical terms and
propose a solution to the problem within the framework of statistical
tests. The proposed solution avoids problems of the currently used
procedures.

\end{abstract}


\maketitle

\section{Introduction}

When existence of a new phenomenon is proposed an experiment is designed 
which exploits the differences between the adopted (old) and the new 
theories to check if there is evidence to reject the old theory in favor 
of the new one. It is this difference which provides the \textit{signal} 
in the experiment. If such a signal is found, a discovery is claimed and 
the values of the parameters of the new theory are measured. If, on the 
other hand, no evidence contradicting the old theory is found, it is 
desirable to set a constraint on the possible values of the parameters 
of the new theory. The logic behind this is simple: if the values of the 
parameters of the new theory were inside a certain region of the 
parameter space, the experiment would have found evidence against the 
old theory in favor of the new one. Since no evidence is actually found, 
such a region is called \textit{the exclusion region}.

Traditionally, the exclusion regions on the parameters of a theory are
constructed based on the upper boundary of a classical one-sided
confidence interval. Often, the exclusion regions constructed by one
experiment rule out signals reported by the others (see, for instance,
CDMS~\cite{cdms} and DAMA~\cite{dama} or LSND~\cite{lsnd} and
KARMEN~\cite{karmen}). Therefore, the task of confidence interval
construction receives considerable attention and is a subject of many 
controversies (see, for
instance,~\cite{FeldmanCousins,fnal,cern,helene99,cdms,dama,lsnd,karmen}). 
There are, however, serious problems associated with the use of
confidence intervals for exclusion region construction which are often
overlooked. Indeed, one of the pre-requisites of the theory of
statistical estimation based on the classical theory of
probabilities~\cite{neyman} is the knowledge that the observed data have
arisen from the phenomenon being observed. In other words, it is
supposed that there is no question whether the observed data $x$ were
drawn from the probability distribution $p_{1}(x;\mu)$; it is known
for a fact and an attempt is being made to quantify the value of the 
parameter $\mu$ by constructing a one-sided confidence interval.

Hence, it is immediately seen that the classical theory of estimation is
not applicable to the situations when it is not known that the
phenomenon exists. The application of the theory to such problems may
lead to intervals which do not have the desired confidence level.
Another problem is that a confidence interval constructed in such a way 
may exclude values of the parameter $\mu$ for which the experiment is
insensitive (see, for instance, discussions
in~\cite{giunti_laveder,FeldmanCousins}).

Thus, the dissatisfaction with the classical theory of estimation is not
due to imperfections in the theory but is caused by misuse of the
theory and by lack of mathematical clarity in problems solutions of
which are sought within the framework of the classical estimation
theory. Yet, it is desirable to be able to construct exclusion regions 
objectively without the use of subjective \textit{priors}.

In this work we formulate the question of what can be stated regarding
the parameter $\mu$ of the hypothesis $H_{1}(\mu)$ of presence of the
new phenomenon when the hypothesis $H_{0}$ of absence of the new
phenomenon is not rejected based on the outcome of the experiment. We
also propose a solution to this problem formulated within the framework
of hypothesis test formalism. In addition, we propose a clear definition
of the \textit{sensitivity} of a detector.


\section{statement of the problem and its solution}

Consider an experiment searching for a new phenomenon where a decision
of plausibility of existence of the new phenomenon is made with the help
of a statistical test~\cite{neyman_pearson}. In such a test, the
hypothesis tested (the null hypothesis $H_{0}$) is the adopted (old)
theory with the alternative hypothesis $H_{1}$ that the observed data is
due to the new phenomenon. Each of the hypotheses defines a probability
distribution of obtaining every possible outcome $x$ of the experiment
$p_{0}(x)$ and $p_{1}(x)$ respectively.  The test is set up so that if
the observed data lie within some critical region $w_{c}$, the null
hypothesis is rejected and is not rejected
otherwise~\cite{neyman_pearson}.

The error leading to an unjust rejection of the null hypothesis is
called the error of the first kind and is denoted by $\alpha$. It is a
common practice to construct the test in such a way as to guarantee that
the error does not exceed a preset value $\alpha_{c}$ called \textit{the
level of significance}. \textit{The power of the test} $(1-\beta)$ is
defined as the probability of rejecting the null hypothesis when the
alternative is true. In other words:

\[ \alpha_{c}\geq\int_{w_{c}}p_{0}(x)dx \ \ \ 
                                     (1-\beta)=\int_{w_{c}}p_{1}(x)dx \]

\subsubsection*{Example}

Suppose that the observable $X$ is distributed according to the Gaussian 
distribution with zero mean and known dispersion $\sigma^{2}$ if the new 
phenomenon does not exist. If, however, the new phenomenon exists, the 
same observable $X$ is distributed according to Gaussian  with the same 
dispersion but positive mean $\mu$. Thus, the hypotheses of the origin 
of the observed data $x$ are :

\[ p_{0}=\frac{1}{\sqrt{2\pi\sigma^{2}}}e^{-x^{2}/2\sigma^{2}} \ \ \ 
   p_{1}=\frac{1}{\sqrt{2\pi\sigma^{2}}}e^{-(x-\mu)^{2}/2\sigma^{2}}
                                                          \ \ \ \mu>0 \]

The best critical region~\cite{neyman_pearson} is defined as $x\geq x_{c}$.
That is, if the observed data point $x$ is greater than $x_{c}$ the null
hypothesis of absence of the new phenomenon is rejected. Thus, in the
proposed test significance and power are:

\[ \alpha_{c}=\frac{1}{\sqrt{2\pi\sigma^{2}}}
                       \int_{x_{c}}^{\infty} e^{-x^{2}/2\sigma^{2}}dx \] 
\[ (1-\beta)=\frac{1}{\sqrt{2\pi\sigma^{2}}}
                 \int_{x_{c}}^{\infty} e^{-(x-\mu)^{2}/2\sigma^{2}}dx \]

The level of significance is often selected at $\alpha_{c}=1.35\cdot 10^{-3}$
which corresponds to $x_{c}=3\sigma$ in this example.

Suppose further that the value of the parameter $\mu$ of the alternative
hypothesis is large (say $\mu=5\sigma$) and the alternative hypothesis
is true. Since the existence of the new phenomenon is reported only if
$x>x_{c}$ is observed, the presence of the new phenomenon will be
established with probability $(1-\beta)=0.997$. If the alternative
hypothesis is true, but the value of the parameter $\mu$ is small (say
$\mu=1\sigma$) the existence of the new phenomenon will be established
only in $0.023$ cases. Thus, it is hopeless to look for the new
phenomenon using the constructed test if the value of $\mu$ is small.

\vspace{1.2\baselineskip}


The general problem which is being addressed in this paper is the
following. Given a critical region $w_{c}$ constructed for a test of a
null hypothesis $H_{0}$ with respect to a composite alternative
hypothesis $H_{1}(\mu)$ with unknown value of $\mu$, what kind of
restriction can be set on admissible values of $\mu$ if, based on the
outcome of the test, the null hypothesis is not rejected. Intuitively,
one would state that the values of the parameter $\mu$ of the
alternative hypothesis for which the null hypothesis can not be rejected
reliably should not be excluded based on the non-rejection of the null
hypothesis.

To formulate this intuitive notion in mathematical terms it should be
realized that the composite hypothesis $H_{1}(\mu)$ can be considered as
a set of simple hypotheses $H_{1}(\mu)$ corresponding to different fixed
values of $\mu$ which can be classified by the power of the test:

\[ (1-\beta(\mu))=\int_{w_{c}}p_{1}(x;\mu)\,dx \]

If the constructed test has low power with respect to the simple
alternative hypothesis $H_{1}(\mu)$, it is not a surprise that no
evidence against the null hypothesis is found. Therefore, if the null
hypothesis is not rejected, the admissible values of the parameter $\mu$
for which the power of the test is small should not be excluded based on
the outcome of the experiment.

If however, the constructed test has a high power with respect to the
simple alternative hypothesis $H_{1}(\mu)$ and no evidence against the
null hypothesis is found, it may be concluded that the admissible values
of the parameter $\mu$ for which the power of the test is higher than
critical value $(1-\beta_{c})$ can be ruled out as unlikely. The
critical value $(1-\beta_{c})$ of the power of the test is motivated by
the problem at hand and should be selected at $90\%$ or higher. The
value $\mu^{c}$ of the parameter $\mu$ corresponding to the smallest
acceptable power of the test $(1-\beta_{c})$ at significance
$\alpha_{c}$ is the \textit{demarcation point} (or \textit{demarcation
hypersurface} if $\mu$ is multi-dimensional) between the allowed and
excluded regions of values of the parameter $\mu$. In the example
considered above, the demarcation point corresponding to the power
$(1-\beta_{c})=0.9$ at significance of $\alpha_{c}=1.35\cdot 10^{-3}$ is
$\mu^{c}=4.3\sigma$. The values of $\mu$ greater than $\mu^{c}$ should
be considered as unlikely when the null hypothesis is not rejected.

Based on the preceding discussion, it is seen that it is the power of
the test which needs to be maximized when constructing experiments.
There are several ways to achieve this. One way to increase the power is
to set a less stringent level of significance (decrease $x_{c}$ in the
example considered) which comes at a price of increased probability to
falsely claim a discovery. The other, perhaps more desirable way to
increase the power is by fundamental modification of the experimental
setup. Such modification can be made keeping the significance level
intact but may increase the operation cost. Examples of this approach
are increased observation time or sample size. In the example considered
here, the decrease in the dispersion $\sigma^{2}$ will increase the
efficacy of the experiment with respect to weak signals keeping the
significance level intact.

\section{Experiment sensitivity}

Another question which needs to be addressed is that of
\textit{sensitivity} of an experiment and what it means. Even though
sensitivity is usually interpreted as the signal strength which a
detector is able to detect, this statement lacks definiteness because
the detection is a statistical process. Due to statistical fluctuations
a strong signal might be missed and a weak signal might be detected.
Thus, the question of sensitivity of the experiment has to be addressed
within the framework of statistical tests as well. Therefore, the
question is: given a critical region $w_{c}$ constructed for a test of a
null hypothesis $H_{0}$ what is the efficacy of the test with respect to
a set of simple alternative hypotheses $H_{1}(\mu)$.

The answer, once again, can be found in terms of significance and power. 
It is reasonable to request that any apparatus to be constructed should
have a chance of signal detection of $50\%$ or more with given level of
significance. That is why it is proposed to quote the sensitivity of a
detector as such signal level that would provide at least $50\%$ power
of the test at the specified level of significance and should not be
regarded as ``absolute'' $100\%$ detection level. In the example
considered above, the sensitivity of the experiment is $\mu=3\sigma$
with significance $1.35\cdot 10^{-3}$.

At this point it is important to note that two identical experiments
looking at identical signal at their sensitivity level may provide
drastically different outcomes. One of the experiments may get lucky and
state a discovery of the phenomenon while the other one may not. It
should be stressed that based on that \textit{it is not} possible to
conclude that outcome of one experiment rules out the signal of the
other one; there is no contradiction between the two. In order to
confirm or refute a signal detection made by an experiment at its
sensitivity level, it is required to conduct a new test which would have
appreciable ($90\%$ or more) power with respect to the claimed signal
strength with pre-specified significance. Returning to the considered
example, to confirm the discovery made on this experiment a new test
would have to be built with the new dispersion
$\sigma^{2}_{new}=0.48\sigma^{2}_{old}$ with the same significance of
$1.35\cdot 10^{-3}$.

\section{Poisson process with known background}

\begin{figure}
\centering
\psfrag{mb}{$\mu_{b}$}
\psfrag{ms}{$\mu_{s}^{c}$}

\includegraphics[width=0.9\columnwidth]{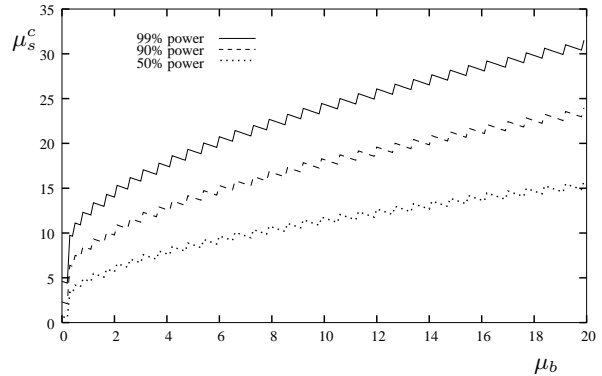}
\caption{\label{fig:known_mu} Signal with average strength 
$\mu_{s}\geq\mu_{s}^{c}$ can be detected with corresponding 
probabilities of at least $50\%$ $90\%$ and $99\%$ at significance 
$1.35\cdot 10^{-3}$ in the presence of known average background 
$\mu_{b}$.}
\end{figure}

\begin{figure}
\centering
\psfrag{mb}{$\mu_{b}$}
\psfrag{ms}{$\mu_{s}^{u}$}

\includegraphics[width=0.9\columnwidth]{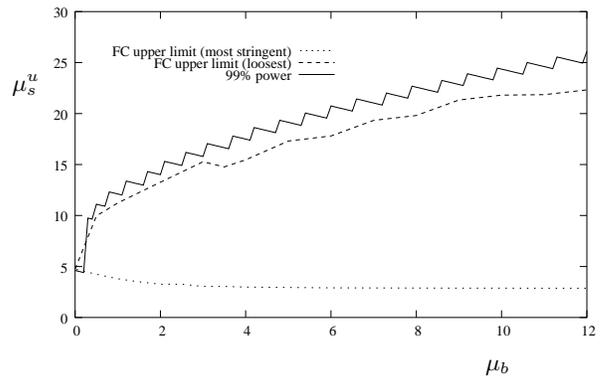}
\caption{\label{fig:compare_mu} 
The upper end $\mu_{s}^{u}$ of the $99\%$ confidence level interval for 
$\mu_{s}$ from tables VIII, IX \cite{FeldmanCousins} can be anywhere 
between the dashed and dotted lines. If the confidence interval were 
used to construct the exclusion region, the signals inside the region 
which are below the solid line could be detected with probability less 
than $99\%$ at significance $1.35\cdot 10^{-3}$.}
\end{figure}

The case of tremendous practical importance is when the number $n$ of
observed events is distributed according to the Poisson distribution

\[ p(n;\mu)=\frac{\mu^{n}}{n!}e^{-\mu} \]

The experiment searching for a new phenomenon may be a subject to 
background so that 

\[ p_{0}(n;\mu_{b})=\frac{\mu_{b}^{n}}{n!}e^{-\mu_{b}} \ \ \ 
          p_{1}(n;\mu_{b}+\mu_{s})=\frac{(\mu_{b}+\mu_{s})^{n}}{n!}
                                               e^{-(\mu_{b}+\mu_{s})} \]
where $\mu_{b}\geq 0$ is the average background rate and $\mu_{s}>0$.

If the average background rate $\mu_{b}$ is known, the best critical 
region against the stated alternative is constructed by

\[ n\geq n_{c} \ \ \ \alpha_{c}\geq\sum_{k=n_{c}}^{\infty}p_{0}(k,\mu_{b})=
                                                    P(n_{c},\mu_{b}) \]
where $P(x,n)$ is complementary regularized incomplete gamma function.

Thus, if no evidence against the null hypothesis is found, the
demarcation point $\mu_{s}^{c}$ on the values of $\mu_{s}$ corresponding
to the power $(1-\beta_{c})$ at significance $\alpha_{c}$ can be
constructed by finding the value of $\mu_{s}^{c}$ such that:

\[ (1-\beta_{c})=\sum_{k=n_{c}}^{\infty}p_{1}(k,\mu_{b}+\mu_{s}^{c})=
                                       P(n_{c},\mu_{b}+\mu_{s}^{c}) \]

Figure~\ref{fig:known_mu} illustrates the situation for the
significance of $1.35\cdot 10^{-3}$ and different requested powers of
the test. It can be seen that even with zero average background events
expected, the signal can be reliably detected (with $99\%$ probability)
only if its average rate is above $\mu_{s}\geq 4.61$. The surprisingly 
high value of the signal is due to discrete nature of the Poisson 
distribution.

It might be interesting to visualize what would happen if the $99\%$ 
confidence interval $[0;\mu_{s}^{u}]$ proposed in~\cite{FeldmanCousins} 
were used for the exclusion region construction. Because the interval 
depends on the number of observed events, the boundary $\mu_{s}^{u}$ may 
be anywhere between the dashed and dotted lines on the 
figure~\ref{fig:compare_mu}. (The dotted line corresponds to the most 
confining interval when zero events is observed while the dashed line 
represents the longest confidence interval obtainable with the left 
boundary fixed to zero. The figure is produced from the tables VIII and 
IX from \cite{FeldmanCousins}.) The values above the boundary 
$\mu_{s}^{u}$ would be excluded with the confidence of $99\%$. It is 
seen that signals inside the exclusion region constructed based on the 
$99\%$ confidence interval~\cite{FeldmanCousins} which are below the 
solid line will be detected with probability much smaller than $99\%$ at 
significance $1.35\cdot 10^{-3}$.

\section{conclusion}

In this report we have considered a problem of what can be stated 
regarding the parameter $\mu$ of alternative hypothesis $H_{1}(\mu)$ of 
presence of a new phenomenon when no evidence against the null 
hypothesis $H_{0}$ of absence of the new phenomenon is found. We have 
proposed a mathematical formulation of this problem and its solution 
within the framework of hypothesis tests theory~\cite{neyman_pearson}.
We have also given reasons why the classical theory of 
estimation~\cite{neyman} is not applicable in situations when the origin 
of data is questioned.

Nevertheless, we recommend to continue to report the classical
confidence intervals assuming that the sought for new phenomenon exists
for at least two reasons. First, the confidence intervals constructed
now may be validated by a future experiment which will discover the
existence of the new phenomenon. Second, the classical confidence 
interval provides information to future experiments about what the value 
of the parameter might be.

However, we propose to discontinue the use of the classical confidence
intervals for construction of exclusion regions when no evidence against
the hypothesis of absence of the new phenomenon is found. Instead, we
propose to construct the exclusion regions based on the power of the
test, since if the undiscovered process existed with the parameter inside
the exclusion region it would have been discovered with probability
$(1-\beta_{c})$ or higher at significance $\alpha_{c}$. Other attractive
features of the constructed exclusion region are that less powerful
experiments will produce less confining exclusion regions, the exclusion
regions do not shrink if the number of observed events is less than the
average expected background; the procedure for exclusion region
construction avoids problems at physical boundaries on the parameter
values and does not exclude the values of the parameter for which the
experiment is insensitive. Also, the procedure of the exclusion region
construction outlined in this paper resolves the illusory contradiction
between the opposite results of two independent observations made at the
sensitivity level of a detector.

It is proposed to call the detector sensitive if at the specified level
of significance at least $50\%$ power of the test can be achieved.


\begin{acknowledgments}

This work is supported by the National Science Foundation (Grant Numbers
PHY-9901496 and PHY-0206656), the U. S. Department of Energy Office of
High Energy Physics, the Los Alamos National Laboratory LDRD program and
the Los Alamos National Laboratory Institute of Nuclear and Particle
Astrophysics and Cosmology of the University of California.

\end{acknowledgments}

\bibliography{upper_limit}
\bibliographystyle{plain}



\end{document}